\documentclass[twocolumn,prl,showkeys]{revtex4}%
\usepackage{graphicx}%
\usepackage{amsmath}%
\usepackage{amsfonts}%
\usepackage{amssymb}
\usepackage{bm}

\def\s{{\sigma}}

\def\k{{ {\bm k} }}
\def\p{{ {\bm p} }}
\def\q{{ {\bm q} }}
\def\Q{{ {\bm Q} }}

\def\w{{\omega}}
\def\a{{\alpha}}
\def\b{{\beta}}

\begin{document}
\title{
Spin-Fluctuation-Driven Orbital Nematic Order in Ru-Oxides:\\
Self-Consistent Vertex Correction Analysis for Two-Orbital Model
}
\author{Yusuke \textsc{Ohno}$^{1}$,
Masahisa \textsc{Tsuchiizu}$^{1}$, 
Seiichiro \textsc{Onari}$^{2}$, and
Hiroshi \textsc{Kontani}$^{1}$
}
\date{\today }

\begin{abstract}
To reveal the origin of the ``nematic electronic fluid phase''
in Sr$_3$Ru$_2$O$_7$, 
we apply the self-consistent vertex correction analysis 
to the ($d_{xz},d_{yz}$)-orbital Hubbard model.
It is found that the Aslamazov-Larkin type vertex correction
causes the strong coupling 
between spin and orbital fluctuations,
which corresponds to the Kugel-Khomskii 
spin-orbital coupling in the local picture.
Due to this mechanism,
orbital nematic order with $C_2$ symmetry is induced by
the magnetic quantum criticality in multiorbital systems,
whereas this mechanism is ignored in the random-phase-approximation.
The present study naturally explains
the intimate relation between 
the magnetic quantum criticality and the nematic state in Sr$_3$Ru$_2$O$_7$
and Fe-based superconductors.
\end{abstract}

\address{
$^1$ Department of Physics, Nagoya University,
Furo-cho, Nagoya 464-8602, Japan. \\
$^2$ Department of Applied Physics, Nagoya University,
Furo-cho, Nagoya 464-8603, Japan. 
}
 
\keywords{Sr$_3$Ru$_2$O$_7$, orbital nematic order, vertex correction,
quantum criticality}

\sloppy

\maketitle

Recently, emergence of orbital (or quadrupole) order
or orbital fluctuations
in multiorbital systems has been attracting great attention.
In heavy-fermion systems, CeB$_6$ exhibits non-magnetic 
quadrupole order \cite{CeB6-Shiba},
and the hidden-order phase in URu$_2$Si$_2$ \cite{Amitsuka,Matsuda-U}
is expected as quadrupole or higher-rank multipole order.
As for $d$-electron systems,
Fe-based superconductors exhibit
``non-magnetic'' orthorhombic structure transition at $T=T_S$
as well as the nematic order, which indicates 
the occurrence of the ferro-orbital polarization $n_{xz}\ne n_{yz}$
 \cite{Shen,Kasahara}.
Large softening of shear modulus $C_{66}$ above $T_S$
indicates the existence of strong orbital fluctuations 
 \cite{Fer,Yoshizawa,Goto}.

The quantum critical phenomenon in bilayer perovskite Sr$_3$Ru$_2$O$_7$
is very unique in that both spin and charge degrees of freedom are 
intimately related
 \cite{Maeno1,Maeno2,Kitagawa,Mac-Rev}.
The band structure of Sr$_3$Ru$_2$O$_7$ is composed of
the $t_{2g}$-orbital ($d_{xz}$, $d_{yz}$, $d_{xy}$ orbitals) of Ru ions.
Both the $T$-linear resistivity \cite{Maeno1} and the NMR result
$1/T_1T\propto T^{-1}$ \cite{Kitagawa} are observed below 20 K
under the magnetic field $H_c\approx 7.8$ Tesla.
These non-Fermi liquid behaviors indicate the emergence of 
``antiferromagnetic quantum criticality (AFM-QC)'' at $H\approx H_c$,
although no long-range magnetic order is observed till 0.1 K
 \cite{Maeno2,Kitagawa}.

Under the critical field $H\approx H_c$, moreover,
Sr$_3$Ru$_2$O$_7$ exhibits a novel ``non-magnetic
nematic electronic fluid phase'' below $1$ K,
which is confirmed by the large anisotropy of in-plane resistivity
 \cite{Maeno2}.
As a possible origin,
the Pomeranchuk instability of the Fermi surfaces (FSs)
in the single-band Hubbard model had been studied using the 
renormalization group method
 \cite{RG1,RG2}
and the perturbation theory
 \cite{Miyake}.
Also, the orbital polarized state has been studied
based on multiorbital Hubbard models,
using the mean-field-level approximation (MFA)
 \cite{Kivelson-MF,Lee-MF,Phillips}.
However, many of these studies do not prove the stability of 
the nematic order against antiferro-spin/orbital order
driven by the nesting of the FS.
Especially, the key question --- why the non-magnetic nematic order 
occurs only near the AFM-QC in Sr$_3$Ru$_2$O$_7$ --- 
has been still unclear.

The nematic order in Fe-based superconductors 
also attracts increasing attention, and it 
would offer us useful hints in the study of Sr$_3$Ru$_2$O$_7$.
Since it cannot be explained in the MFA analysis,
the spin nematic order due to the order-by-disorder
mechanism had been proposed \cite{Fer,Kivelson}.
However, it could not be applied to Sr$_3$Ru$_2$O$_7$
since the incommensurate spin fluctuations are realized.
On the other hand, we have recently revealed that 
the vertex correction (VC), which describes the many-body effect
beyond the MFA, induces strong ferro-orbital fluctuations
 \cite{Onari-SCVC,Kontani-Rev}.
Then, it is highly required to analyze the significance of the VC in Ru-oxides.

In this paper, we study the origin of nematic phase
in Sr$_3$Ru$_2$O$_7$ based on the two-orbital Hubbard Hamiltonian.
We utilize the self-consistent VC (SC-VC) method, 
which was recently applied to Fe-based superconductors successfully,
and reveal that the ``orbital Pomeranchuk instability''
is generally induced near AFM-QC,
owing to the spin-orbital coupling given by the VC.
The present study predicts the realization of orbital nematic order
near the field-induced AFM-QC in Sr$_3$Ru$_2$O$_7$.

\begin{figure}[!htb]
\includegraphics[width=.99\linewidth]{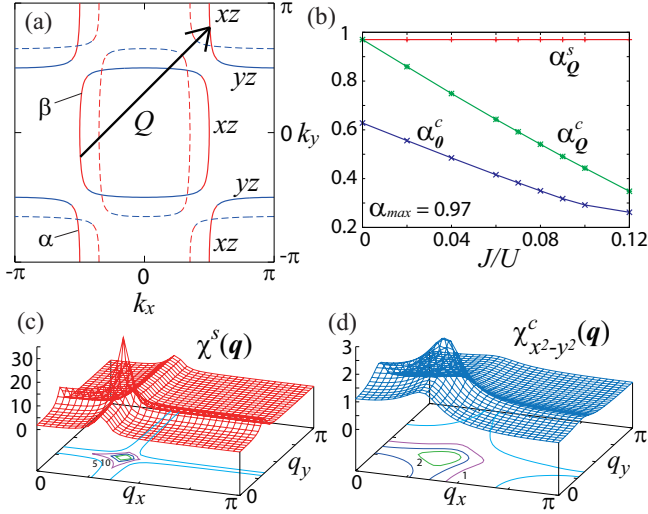}
\caption{(color online)
(a) FSs of the two-orbital model for $n=2$.
The colors correspond to $1=xz$ (red) and $2=yz$ (blue).
The arrow represents the $xz$ intra-orbital nesting vector.
The dashed lines represent the deformed FSs for 
$\langle {\hat O}_{x^2-y^2}\rangle<0$.
(b) $\a^s_{\bm Q}$, $\a^c_{\bm Q}$ and $\a^c_{\bm 0}$ for $n=3.3$ in the RPA.
(c) $\chi^s(\q)$ and (d) $\chi^q(\q)$ obtained by the RPA.
}
\label{fig:fig1}
\end{figure}

The bandstructure of Sr$_3$Ru$_2$O$_7$ is rather complex
because of the tilting of RuO$_6$ octahedra.
Therefore, we study the simplified ($d_{xz},d_{yz}$)-orbital model
to grasp the essential mechanism of the orbital nematicity:
\begin{eqnarray}
H&=&\sum_{\k;\s=\uparrow,\downarrow;\mu,\nu=1,2} 
\xi_\k^{\mu\nu} c_{\k,\mu,\s}^\dagger c_{\k,\nu,\s} +H_{\rm C},
\label{eqn:Ham}
\end{eqnarray}
where $\mu,\nu=1,2$ represents the $d$-orbital; $1=xz$ and $2=yz$.
This model describes the $\a$ FS and $\b$ FS
of Ru-oxides, and
it was analyzed in the study of 
anomalous/spin Hall effect \cite{Kontani-AHE}.
Hereafter,
we promise that $x,y$-axes are along the nearest Ru-Ru bond directions.
Then, the intra- and inter-orbital hoppings are given as
$\xi_\k^{11}=-2t\cos k_x$, 
$\xi_\k^{22}=-2t\cos k_y$, and
$\xi_\k^{12}= 4t'\sin k_x \sin k_y$.
In this paper, we put $(t,t')=(1,0.1)$.
The FSs for the electron filling $n=2$ are shown in Fig. \ref{fig:fig1} (a).
In Sr$_3$Ru$_2$O$_7$, ($\a,\b$) FSs split into the bonding ($\a_1,\b_1$) FSs 
and the antibonding ($\a_2,\b_2$) FSs by large interlayer hoppings. 
The electron filling of ($\a_1,\b_1$) FSs is about $3.2-3.5$ \cite{Kivelson-MF,Lee-MF}, 
and we set $n=3.3$ in this study.
$H_{\rm C}$ represents the multiorbital Coulomb
interaction composed of intra (inter) orbital interaction 
$U$ ($U'$) and the exchange interaction $J$.
\cite{Kontani-Rev}.
Hereafter, we put $U=U'+2J$.

In the present model, the susceptibility for the charge (spin)
channel is given by the following $4\times4$ matrix form 
in the orbital basis:
\begin{eqnarray}
{\hat \chi}^{c(s)}(q)={\hat {\chi}}^{{\rm irr},c(s)}(q)
 (1-{\hat \Gamma}^{c(s)}{\hat {\chi}}^{{\rm irr},c(s)}(q))^{-1} ,
\label{eqn:chi}
\end{eqnarray}
where $q=(\q,\w_l=2\pi l T)$, and
${\hat \Gamma}^{c(s)}$ represents the Coulomb interaction
for the charge (spin) channel 
composed of $U$, $U'$ and $J$ given in Refs. \cite{Kontani-Rev}.
The irreducible susceptibility is 
\begin{eqnarray}
{\hat {\chi}}^{{\rm irr},c(s)}(q)= {\hat \chi}^{0}(q)+{\hat X}^{c(s)}(q) ,
\label{eqn:irr}
\end{eqnarray}
where $\chi^{0}_{ll',mm'}(q)=-T\sum_p G_{lm}(p+q)G_{m'l'}(p)$ is the bare bubble,
$p=(\p,\epsilon_n=(2n+1)\pi T)$,
and the second term is the VC that is neglected in the RPA.

In the present discussion, it is convenient to consider the 
quadrupole susceptibilities:
$\chi^c_{\gamma}(\q)\equiv
\sum_{ll',mm'} O_\gamma^{l,l'}{\chi}_{ll',mm'}^c(\q)O_{\gamma}^{m',m}$,
where $O_\gamma^{l,m}=\langle l|{\hat O}_\gamma |m\rangle$
is the matrix element of the $\gamma$-quadrupole operator.
Non-zero matrix elements of the quadrupole operators are
$O_{x^2-y^2}^{1,1}=-O_{x^2-y^2}^{2,2}=1$ and
$O_{xy}^{1,2}=O_{xy}^{2,1}=1$
 \cite{Kontani-Rev},
while ${\hat O}_{xz,yz}={\hat 0}$ in the present Hilbert space.

The divergence of $\chi^c_{x^2-y^2}({\bm 0})$ immediately leads to the  
ferro-quadrupole order $\langle {\hat O}_{x^2-y^2}\rangle=n_{xz}-n_{yz}\ne0$,
resulting in the ``nematic'' deformation of the FSs shown 
in the dashed lines in Fig. \ref{fig:fig1} (a).
The director of the nematicity is along the Ru-Ru bond direction,
which is consistent with experiments.
In case of $\langle {\hat O}_{xy}\rangle\ne0$,
the director is rotated by $45^\circ$.

Here, we introduce the charge (spin) Stoner factor $\a^{c(s)}_\q$, 
which is the largest eigenvalue of 
${\hat \Gamma}^{c(s)}{\hat {\chi}}^{{\rm irr},c(s)}(\q)$ at $\w_l=0$:
Then, the quadrupole (spin) susceptibility is enhanced
in proportion to $(1-\a_{\rm max}^{c(s)})^{-1}$, where
$\a_{\rm max}^{c(s)}\equiv {\rm max}_\q\{\a^{c(s)}_\q\}$.
Since $J>0$ in real systems, spin fluctuations are always dominant 
($\a^s_{\rm max}>\a^c_{\rm max}$) in the RPA as shown in Fig. \ref{fig:fig1} (b):
This figure shows the $J/U$-dependences of $\a^s_{\bm Q}$, $\a^c_{\bm Q}$ and $\a^c_{\bm 0}$ 
in the RPA, where $U$ is determined by the condition $\a^s_{\bm Q}(U)=0.97$.
In the SC-VC method, however, 
the opposite relation $\a^s_{\rm max}\lesssim\a^c_{\rm max}$ can be realized 
even for $J/U\lesssim 0.1$ 
because of large ${\hat X}^c(\q)$.

First, we perform the RPA calculation for $n=3.3$ and $T=0.05$,
using $64\times64$ $\k$-meshes.
The unit of energy is $t=1$.
Figure \ref{fig:fig1} (c) shows the spin susceptibility
$\chi^s(\q)=\sum_{l,m}\chi^s_{ll,mm}(\q)$ and the 
(d) quadrupole susceptibility $\chi^c_{x^2-y^2}(\q)$ for $J/U=0.06$.
The Stoner factors are 
$\a^s_{\rm max}=0.97$, $\a^c_\Q=0.64$, and $\a^c_{\bm 0}=0.42$;
see Fig. \ref{fig:fig1} (b).
Both $\chi^s(\q)$ and $\chi^c_{x^2-y^2}(\q)$ have peaks 
at $(q^*,q^*)$, where $q^*=0.31\pi$.
Thus, the RPA cannot explain the nematic transition
that requires the divergence of $\chi^c_{x^2-y^2}({\bm 0})$.

In the next stage, we study the role of VC
due to the Maki-Thompson (MT) and Aslamazov-Larkin (AL) terms
shown in Fig. 2 (a) of Ref. \cite{Onari-SCVC}.
They are given by the Ward identity 
$\Gamma^I=\delta\Sigma_{\rm FLEX}/\delta G$ 
using the FLEX self-energy.
Moreover, they correspond to the first-order
mode-coupling corrections to the RPA susceptibility:
The intra- (inter-) bubble correction gives the MT (AL) term
\cite{Moriya}.
In single-orbital models, these VCs
had been studied by the self-consistent-renormalization (SCR)
theory \cite{Moriya}. 
However, significant role of the AL-type VC in multiorbital systems
had been overlooked until recently \cite{Onari-SCVC}.

The charge AL term ${\hat X}^{{\rm AL},c}(q)\equiv 
{\hat X}^{{\rm AL},\uparrow,\uparrow}(q)+{\hat X}^{{\rm AL},\uparrow,\downarrow}(q)$ 
is given in eq. (5) of Ref. \cite{Onari-SCVC}.
Also, the $ll',mm'$ component of the
spin AL term ${\hat X}^{{\rm AL},s}(q)\equiv 
{\hat X}^{{\rm AL},\uparrow,\uparrow}(q)-{\hat X}^{{\rm AL},\uparrow,\downarrow}(q)$ 
is given as
\begin{eqnarray}
& &\frac{T}2\sum_{k}\sum_{a\sim h}
\Lambda_{ll',ab,ef}(q;k)[ \{  {V}_{ab,cd}^s(k+q){V}_{ef,gh}^c(-k)
\nonumber \\
& &\ \ +{V}_{ab,cd}^c(k+q){V}_{ef,gh}^s(-k) \}
\Lambda_{mm',cd,gh}'(q;k) 
 \nonumber \\
& &\ \ \ +2{V}_{ab,cd}^s(k+q){V}_{ef,gh}^s(-k)\Lambda_{mm',cd,gh}''(q;k) ],
 \label{eqn:ALexample2}
\end{eqnarray}
where 
${\hat V}^{s,c}(q)\equiv{\hat \Gamma}^{s,c}
+ {\hat\Gamma}^{s,c}{\hat\chi}^{s,c}(q){\hat\Gamma}^{s,c}$.
${\hat \Lambda}(q;k)$ is the three-point vertex \cite{Onari-SCVC}, and
$\Lambda_{mm',cd,gh}'(q;k)\equiv
\Lambda_{ch,mg,dm'}(q;k)+\Lambda_{gd,mc,hm'}(q;-k-q)$ and
$\Lambda_{mm',cd,gh}''(q;k) \equiv
\Lambda_{ch,mg,dm'}(q;k)-\Lambda_{gd,mc,hm'}(q;-k-q)$.
Since ${\hat X}^{{\rm AL},c}({\bm 0})\sim \sum_k{\hat \Lambda}
(3{\hat V}^{s}(k)^2+{\hat V}^{c}(k)^2){\hat \Lambda}'$, 
it becomes significant
when either ${\hat \chi}^c$ or ${\hat \chi}^s$ is large,
while ${\hat X}^{{\rm AL},s}$ is less important unless 
both ${\hat \chi}^c$ and ${\hat \chi}^s$ are large
since ${\hat \Lambda}''\ll{\hat \Lambda}'$.
For this reason, in Ref. \cite{Onari-SCVC},
we have calculated only ${\hat X}^{c}(\q)$
by putting ${\hat X}^{s}(\q)=0$ for simplicity.
In the present study, we also calculate both 
${\hat X}^{c}(\q)$ and ${\hat X}^{s}(\q)$ self-consistently.
Hereafter, we call the former (latter) the SC-VC[c] (SC-VC[all]) method.
We will show that both methods give similar results.

\begin{figure}[!htb]
\includegraphics[width=.99\linewidth]{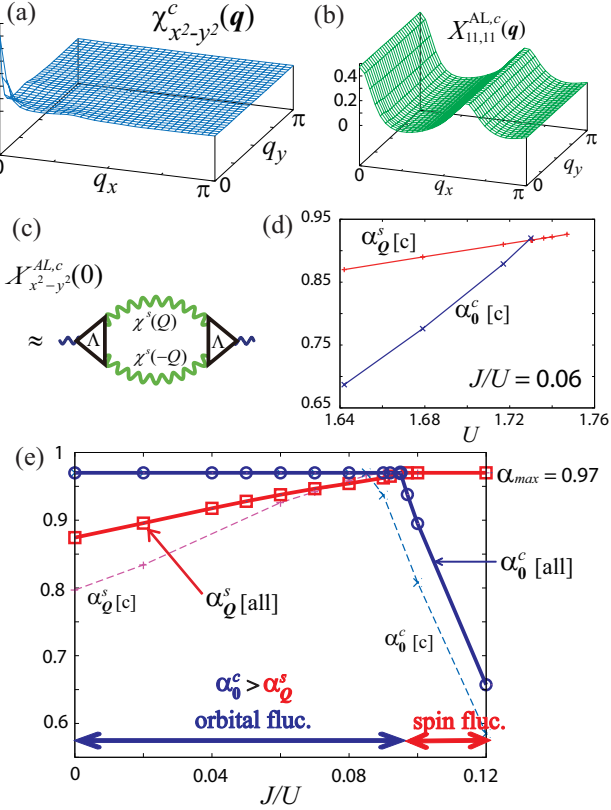}
\caption{(color online)
(a) $\chi^c_{x^2-y^2}(\q)$ and (b) $X^{{\rm AL},c}_{11,11}(\q)$
given by the SC-VC[c] method for $U=1.75$ and $J/U=0.06$.
(c) the charge AL term.
(d) $\a^{s}_\Q$ and $\a^{c}_{\bm 0}$ in the SC-VC[c] method
as function of $U$ under the condition $J/U=0.06$.
(e) $\a^{s}_\Q$ and $\a^{c}_{\bm 0}$ as function of $J/U$
in the SC-VC[all] and SC-VC[c] methods, 
under the condition $\a_{\rm max}\equiv{\rm max}\{\a^s_{\rm max},\a^c_{\rm max}\}=0.97$.
}
\label{fig:fig2}
\end{figure}

First, we present the numerical results 
given by the SC-VC[c] method:
Figure \ref{fig:fig2} (a) shows $\chi^c_{x^2-y^2}(\q)$ obtained
for $n=3.3$, $J/U=0.06$ and $U=1.75$, in which the Stoner factors are
$\a^s_{\rm max}=0.93$ and $\a^c_{\bm 0}=0.97$.
In comparison with the RPA, $\chi^c_{x^2-y^2}(\q)$ is strongly enhanced 
by the charge AL term, while the results are almost unchanged 
even if MT term is dropped.
Figure \ref{fig:fig2} (b) presents the 
momentum dependence of the charge VC, $X_{11,11}^{{\rm AL},c}(\q)$,
which is enhanced on lines $q_x=0, 2q^*$ because of the 
two-magnon process in Fig. \ref{fig:fig2} (c), reflecting the
good nesting of the quasi one-dimensional $xz$-orbital bands.
(Fig. \ref{fig:fig2} (c) shows a virtual decay process of 
an orbiton into two magnons with opposite momenta $\pm \Q$.)
In the same way, $X_{22,22}^c(\q)$ is enhanced on lines $q_y=0,2q^*$.
Thus, both $X_{11,11}^c(\q)$ and  $X_{22,22}^c(\q)$
give the enhancement of $\chi^c_{x^2-y^2}(\q)$ at $\q={\bm 0}$.
In contrast, $\chi^c_{xy}(\q)$ is enhanced only slightly
since the three point vertex $\Lambda$ for $O_{xy}$-quadrupole
is much smaller.
Figure \ref{fig:fig2} (d) gives the $U$-dependence 
of $\a^c_{\bm 0}$ and $\a^s_{\bm Q}$ under the constraint $J/U=0.06$
in the SC-VC[c] method.
Since the slope of $\a^c_{\bm 0}$ is larger than that of $\a^s_{\bm Q}$,
the critical value $(J/U)_c$ increases in the strong correlation region.

Now, we discuss the mechanism of the enhancement of 
$\chi^c_{x^2-y^2}({\bm 0})$ in more detail:
Considering the small inter-orbital mixing due to $t'(\ll t)$,
we drop $\chi_{ll',mm'}^{{\rm irr},c}({\bm 0})$ in eq. (\ref{eqn:chi}) 
except for $l=l'=m=m'$.
Then, eq. (\ref{eqn:chi}) is simplified to $2\times2$ matrix equation with
\begin{eqnarray}
{\hat \chi}^c({\bm 0})&=&
\left(
\begin{array}{cc}
\chi_{11,11}^c({\bm 0}) & \chi_{11,22}^c({\bm 0}) \\
\chi_{11,22}^c({\bm 0}) & \chi_{11,11}^c({\bm 0}) \\
\end{array}
\right),
 \nonumber \\
{\hat \chi}^{{\rm irr},c}({\bm 0})&=&
\left(
\begin{array}{cc}
\chi_{11,11}^{{\rm irr},c}({\bm 0}) & 0 \\
0 & \chi_{11,11}^{{\rm irr},c}({\bm 0}) \\
\end{array}
\right),
 \nonumber \\
{\hat \Gamma}^{c}&=&-
\left(
\begin{array}{cc}
U & 2U'-J \\
2U'-J & U \\
\end{array}
\right).
\end{eqnarray}
For $U'=U-2J$,
the charge density susceptibility 
$\chi^c_{n}({\bm 0})= \sum_{l,m}\chi_c^{ll,mm}({\bm 0})$
and $\chi^c_{x^2-y^2}({\bm 0})$ are obtained as
\begin{eqnarray}
\chi^c_{n}({\bm 0})&=& 2\chi_{11,11}^{{\rm irr},c}({\bm 0})
(1+(3U-5J)\chi_{11,11}^{{\rm irr},c}({\bm 0}))^{-1}
 \label{eqn:chic} \\
\chi^c_{x^2-y^2}({\bm 0})&=&2\chi_{11,11}^{{\rm irr},c}({\bm 0})
(1-(U-5J)\chi_{11,11}^{{\rm irr},c}({\bm 0}))^{-1}
\label{eqn:chiq}
\end{eqnarray}
Therefore, $\chi^c_{x^2-y^2}({\bm 0})$ is enhanced
by Coulomb interaction when $J/U<0.2$,
while $\chi^c_{n}({\bm 0})$ is always suppressed.
If we drop the spin VC, the spin susceptibility is 
$\chi^s(\Q)=2\chi_{11,11}^{0}(\Q)(1-(U+J)\chi_{11,11}^{0}(\Q))^{-1}$.
Therefore, the relation $\a^c_{\bm 0}>\a^s_{\bm Q}$ 
holds for $X_{11,11}^{c}({\bm 0})>(U+J)\chi_{11,11}^{0}(\Q)(U-5J)^{-1}
-\chi_{11,11}^{0}({\bm 0})$ for $J/U<0.2$.
Since $X^{{\rm AL},c}({\bm 0})$ grows in proportion to $T\chi^s(\Q)$ 
[$\log \{\chi^s(\Q)\}^2$] at high [low] temperatures \cite{Onari-SCVC},
$\chi^c_{x^2-y^2}({\bm 0})$ is strongly enhanced near the AFM-QC 
under the condition $J/U\lesssim0.2$.
This condition is expected to be satisfied in Ru-oxides.

Figure \ref{fig:fig2} (e) shows the $J/U$-dependence
of $\a^c_{\bm 0}$ and $\a^s_{\bm Q}$ given in the SC-VC[all] 
and SC-VC[c] methods, by adjusting $U$ to satisfy 
$\a_{\rm max}\equiv{\rm max}\{\a^s_{\rm max},\a^c_{\rm max}\}=0.97$.
In both methods, the obtained results are similar
since ${\hat X}^s$ is less important.
The critical value of $J/U$,
at which $\a^c_{\bm 0}=\a^s_{\bm Q}=0.97$ is satisfied,
is $(J/U)_c=0.095 \ (0.086)$ in the SC-VC[all] (SC-VC[c]) method.
Therefore, the relation $\a^c_{\bm 0}>\a^s_{\bm Q}$
is satisfied for $J/U<(J/U)_c\sim0.1$ in the present SC-VC method,
in highly contrast to the PRA result $(J/U)_c^{\rm RPA}=0$;
see Fig. \ref{fig:fig1} (b).

\begin{figure}[!htb]
\includegraphics[width=.99\linewidth]{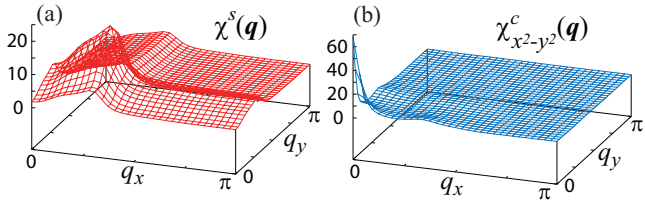}
\caption{(color online)
(a) $\chi^s(\q)$ and (b) $\chi^c_{x^2-y^2}(\q)$ 
obtained by the SC-VC[all] method for $U=1.35$ and $J/U=0.06$.
}
\label{fig:fig3}
\end{figure}

Figure \ref{fig:fig3} shows (a) $\chi^s(\q)$ and (b) $\chi^c_{x^2-y^2}(\q)$
given by the SC-VC[all] method for $U=1.35$ and $J/U=0.06$.
The Stoner factors are $\a^s_{\bm Q}=0.94$ and $\a^c_{\bm 0}=0.97$.
In this method, the shape of $\chi^s(\q)$ is 
slightly changed from the RPA result
by the momentum dependence of ${\hat X}^s(\q)$.
However, the overall results of spin and 
quadrupole susceptibilities are unchanged by the spin VC.
The obtained divergent behavior of $\chi^c_{x^2-y^2}({\bm 0})$
is consistent with the ``non-magnetic nematic phase'' 
in Sr$_3$Ru$_2$O$_7$ near the field-induced AFM-QC.
Recent ultrasonic measurement of the elastic constants
reports a large softening of shear modulus 
$C_S=(C_{11}-C_{12})/2$ under $H_z\sim7.8$ Tesla
\cite{Suzuki}.
Since $C_S^{-1}-C_{S,0}^{-1}\propto \chi^c_{x^2-y^2}({\bm 0})$
($C_{S,0}$ being the lattice shear modulus),
the obtained development of $\chi^c_{x^2-y^2}({\bm 0})$
in Fig. \ref{fig:fig3} (b) is consistent with experiment. 
We have verified that obtained results in Fig. \ref{fig:fig2} and \ref{fig:fig3} 
are qualitatively unchanged for $3.1 \leq n \leq 3.6$.
We will report detailed $n$ dependence of $\chi^c_{x^2-y^2}$ and $\chi^s$ in later publication.

The role of the AL term can be interpreted as the 
following effective nematic interaction
within the mean-field picture:
$H_{\rm nem}\sim g\sum_{\k,\k'}{\hat O}_{x^2-y^2}(\k){\hat O}_{x^2-y^2}(\k')$, 
where ${\hat O}_{x^2-y^2}(\k)= \sum_{l,m,\s} O_{x^2-y^2}^{l,m}c_{\k,l,\s}^\dagger c_{\k,m,\s}$
and $g$ is approximately given as
$(-U+5J)(X^c_{11,11}(\q)+X^c_{22,22}(\q))/2\chi^0(\q)$ at $\q={\bm 0}$.
according to eq. (\ref{eqn:chiq}).
In previous studies of Sr$_3$Ru$_2$O$_7$ based on 
the single-band model, on the other hand, 
the phenomenological Pomeranchuk interaction 
$H_{\rm nem}'\sim g'\sum_{\k,\k'} d_\k d_{\k'}$
had been frequently introduced,
where $d_\k=(\cos k_x -\cos k_y)\sum_\s c_{\k,\s}^\dagger c_{\k,\s}$.
They are similar in symmetry, since
both $d_\k$ and ${\hat O}_{x^2-y^2}(\k)$
belong to the same $B_{1g}$ representation of $D_{4h}$ point group.
Thus, ``orbital Pomeranchuk interaction'' is driven by
the AL-VC in multiorbital models.

It is useful to comment on the difference between the present study
and the study of the five-orbital model in Ref. 
\cite{Onari-SCVC}.
In both models, we obtain the development of 
$\chi^c_{x^2-y^2}({\bm 0})$ using the SC-VC method.
In the latter model, in addition, 
$\chi^c_{xz(yz)}({\bm Q})$ also develops 
because of the VC and good {\it inter-orbital} ($xz/yz$-$xy$) nesting.
Both ferro- and antiferro-quadrupole fluctuations 
induce the $s$-wave superconductivity
without sign reversal ($s_{++}$-wave state)
\cite{Onari-SCVC}.
On the other hand, $\chi^c_{xz(yz)}(\q)$ will be small
in the $t_{2g}$-orbital model for Ru-oxides
because of ill {\it inter-orbital} nesting.
Thus, the results of this paper
would be qualitatively unchanged in the three-orbital model.

Finally, we discuss the physical meaning of the VC 
in the strong-coupling regime ($U\gg t$):
Due to the Kugel-Khomskii type spin-orbital exchange coupling 
$\sim (t^2/U)({\bm s}_i\cdot{\bm s}_j)(O_{x^2-y^2}^i\cdot O_{x^2-y^2}^j)$,
the antiferro-spin order induces the ferro-orbital order, and vise versa.
Such spin-orbital coupling should exist also in the metallic state
($U\sim t$), and it is actually described by the AL-type VC.
For this reason, cooperative development of spin 
and orbital fluctuations is obtained in the SC-VC analysis.
[The RPA is insufficient in multiorbital Hubbard models
in that the spin-orbital coupling is completely ignored.]
Especially, non-magnetic orbital nematic order can be realized
since the scalar order parameter $O_{x^2-y^2}$
is more stable than the vector order parameter ${\bm s}$ 
against the quantum and thermal fluctuations.

In summary, we have studied the origin of non-magnetic
nematic order in Sr$_3$Ru$_2$O$_7$, 
by applying the SC-VC method to the two-orbital Hubbard model.
We have found that the present model
exhibits the orbital Pomeranchuk instability
near the magnetic quantum criticality,
owing to the spin-orbital coupling described by the VC.
For $J/U< (J/U)_c\sim0.1$,
the ferro-orbital order ($\a^c_{\bm 0}\approx1$)
occurs prior to the magnetic transition ($\a^s_\Q\approx1$).
The present mechanism gives a natural explanation for the 
nematic order in Sr$_3$Ru$_2$O$_7$ as well as 
Fe-based superconductors near AFM-QC, 
and it will be also realized in various multiorbital models.
As the origin of the field-induced AFM-QC.
Both the van-Hove singularity \cite{Yamase-vHS,Sigrist}
and the field-suppression of quantum fluctuation \cite{Sakurazawa}
mechanisms had been discussed.

We note that the renormalization group (RG) method
is very powerful for the study of VCs in low dimensional systems.
Recently, the RG analysis has been performed 
for the ($d_{xz}$,$d_{yz}$)-orbital Hubbard model
\cite{Tsuchiizu},
and revealed that $\chi^c_{x^2-y^2}({\bm 0})$ is critically enhanced by
both the magnetic and superconducting QCs.

\acknowledgements
This study has been supported by Grants-in-Aid for Scientific 
Research from MEXT of Japan.
Part of numerical calculations were
performed on the Yukawa Institute Computer Facility.


\end{document}